\documentclass{article}

\PassOptionsToPackage{numbers, sort&compress}{natbib}

\usepackage[final]{neurips_2023_ml4ps}

\usepackage[utf8]{inputenc} 
\usepackage[T1]{fontenc}    
\usepackage[colorlinks, citecolor=LimeGreen!66!Green, urlcolor=Blue]{hyperref}       
\usepackage{url}            
\usepackage{booktabs}       
\usepackage{amsfonts}       
\usepackage{nicefrac}       
\usepackage{microtype}      
\usepackage[svgnames]{xcolor}         

\usepackage{bm}
\usepackage{xspace}
\usepackage{physics}
\usepackage{siunitx}
\usepackage{interval}
\usepackage{cleveref}
\usepackage{tikz}
\usepackage[style=american]{csquotes}
\MakeOuterQuote{"}
\usepackage{enumitem}
\usepackage{wrapfig}
\usepackage{caption}
\usepackage{subcaption}
\usepackage{multicol,multirow}

\defcitealias{Thorp_Mandel_2022}{TM22}

\newcommand{\eg}{e.g.\@\xspace}
\newcommand{\ie}{i.e.\@\xspace}
\newcommand{\cf}{cf.\@\xspace}

\usepackage{glossaries}
\glsdisablehyper

\newacronym{csp}{CSP}{Carnegie Supernova Project}

\newacronym{bhm}{BHM}{Bayesian hierarchical model}

\newacronym{sbi}{SBI}{simulation-based inference}
\newacronym{nre}{NRE}{neural ratio estimation}
\newacronym{tmnre}{TMNRE}{truncated marginal neural ratio estimation}

\newacronym{gpu}{GPU}{graphics processing unit}
\newacronym{nn}{NN}{neural network}
\newacronym{relu}{ReLU}{rectified linear unit}

\newacronym{bce}{BCE}{binary cross-entropy}

\newcommand{\Rnum}[1]{\uppercase\expandafter{\romannumeral#1\relax}}

\edef\aenoxspace{\ae}
\appto{\ae}{\xspace}
\newcommand{\Ianoxspace}{\Rnum{1}a}
\newcommand{\Ia}{\Ianoxspace\xspace}
\newcommand{\SN}{SN\xspace}
\newcommand{\SNae}{SN\aenoxspace\xspace}
\newcommand{\SNIa}{SN~\Ianoxspace\xspace}
\newcommand{\SNaeIa}{SN\aenoxspace~\Ianoxspace\xspace}

\newcommand{\SNANA}{SNANA\xspace}

\NewDocumentCommand{\newmathcommand}{m O{0} m}{\newcommand{#1}[#2]{\ensuremath{#3}\xspace}}
\NewDocumentCommand{\renewmathcommand}{m O{0} m}{\renewcommand{#1}[#2]{\ensuremath{#3}\xspace}}

\newcommand{\newoperator}[3][\operatorname]{
    \newcommand{#2}[1][]{\ensuremath{\opbraces{#1{#3}\ifblank{##1}{}{_{##1}}}}}
}

\newoperator{\Normal}{\mathcal{N}}
\newoperator{\Uniform}{\mathcal{U}}

\newmathcommand{\data}{\mathbf{d}}
\newoperator{\Model}{\mathcal{M}}
\newmathcommand{\params}{\bm{\theta}}

\newoperator{\prob}{p}
\newcommand{\given}{\,|\,}
\newoperator{\Expectation}{\mathds{E}}

\newcommand{\mprior}[1][k]{\ensuremath{\prob(\Model[#1])}\xspace}
\newcommand{\mpost}[1][k]{\ensuremath{\prob(\Model[#1]\given\data)}\xspace}
\newcommand{\evidence}[1][k]{\ensuremath{\prob(\data\given\Model[#1])}\xspace}
\newcommand{\prior}[1][k]{\ensuremath{\prob(\params_{#1})}\xspace}
\newcommand{\likelihood}[1][k]{\ensuremath{\prob(\data\given\params_{#1})}\xspace}
\newcommand{\joint}{\ensuremath{\prob(\params, \data)}\xspace}

\newmathcommand{\nsn}{N_{\text{\SN}}}
\newmathcommand{\nsnobs}{N^s_{\text{obs}}}
\newoperator{\Linear}{Linear}
\newoperator{\WhitenOnline}{WhitenOnline}
\newoperator{\ReLU}{ReLU}
\newoperator{\Dropout}{Dropout}
\newoperator{\LogSoftmax}{LogSoftmax}
\newoperator{\NLLLoss}{NLLLoss}
\newoperator{\CrossEntropyLoss}{CrossEntropyLoss}

\newmathcommand{\stepmag}{\Delta M}
\newmathcommand{\sigmares}{\sigma_0}
\newmathcommand{\meanRv}{\mu_{R}}
\newmathcommand{\sigmaRv}{\sigma_{R}}
\newmathcommand{\Rvs}{R_V^s}

\title{SimSIMS: Simulation-based Supernova Ia Model Selection with thousands of latent variables}

\author{%
    \shortstack{Konstantin Karchev\textsuperscript{1} \\[0.5em] \texttt{kkarchev@sissa.it}}
    \qquad\qquad
    \shortstack{Roberto Trotta\textsuperscript{1,2,3,4} \\[0.5em] \texttt{rtrotta@sissa.it}}
    \qquad\qquad
    \shortstack{Christoph Weniger\textsuperscript{5} \\[0.5em] \texttt{c.weniger@uva.nl}}
    \\[1em]
    \shortstack[l]{
    \textsuperscript{1} Theoretical and Scientific Data Science, SISSA, Trieste, Italy\\
    \textsuperscript{2} Department of Physics, Imperial College London, United Kingdom \\
    \textsuperscript{3} Italian Research Center on High Performance Computing, Big Data and Quantum Computing \\
    \textsuperscript{4} National Institute for Nuclear Physics, Trieste, Italy \\
    \textsuperscript{5} GRAPPA Institute, University of Amsterdam, The Netherlands
    }
}

\renewcommand{\paragraph}[1]{{\bf #1}\hspace{0.5em}}

\begin{document}

\maketitle

\begin{abstract}
  We present principled Bayesian model comparison through simulation-based neural classification applied to \SNIa analysis. We validate our approach on realistically simulated \SNIa light curve data, demonstrating its ability to recover posterior model probabilities while marginalizing over $>\num{4000}$ latent variables. The amortized nature of our technique allows us to explore the dependence of Bayes factors on the true parameters of simulated data, demonstrating Occam's razor for nested models. When applied to a sample of \num{86} low-redshift \SNaeIa from the \acrlong{csp}, our method prefers a model with a single dust law and no magnitude step with host mass, disfavouring different dust laws for low- and high-mass hosts with odds in excess of \num{100}:\num{1}.
\end{abstract}

\section{Introduction}

Classification problems are a quintessential machine learning task, just as hypothesis testing is at the heart of science. Bayesian model selection improves upon traditional frequentist tests by implementing an automatic quantitative version of Occam's razor (the principle that "simple" models ought to be preferred~\cite{Trotta2008}). Traditionally, calculating Bayesian model evidences has required performing an integral over the model's whole parameter space, which quickly becomes intractable when analysing large data sets with complicated \glspl{bhm}.

Neural \gls{sbi}\footnote{See \cite{Cranmer_2020,Lueckmann_2021} for overviews and \url{https://simulation-based-inference.org/} and \url{https://github.com/smsharma/awesome-neural-sbi} for references to software and applications.} is a relatively recent alternative approach to Bayesian inference that is rapidly gaining popularity in the physical sciences due to its scalability to large data sets and ability to include realistic models.
The keystone of \gls{sbi} is the use of a stochastic simulator able to produce mock data, incorporating arbitrarily complex physical effects difficult to model in likelihood-based pipelines. A \gls{nn} trained on the simulated examples is then used for inference in place of explicit likelihood evaluations. \Glspl{nn} are quick to train via gradient descent, easy to deploy on modern high-performance computing hardware like \glspl{gpu}, and allow \gls{sbi} practitioners to exploit the rapid development in the field of deep learning. Furthermore, amortised inference enables both validation of the approximate posteriors \citep{Hermans_2022,Delaunoy_2022,Miller:2021hys} as well as the constructions of confidence regions with guaranteed frequentist coverage \citep{Karchev-sicret,Crisostomi_2023}.

Here, we combine the power of \gls{sbi} with the elegance of Bayesian model selection to perform principled analysis of a \gls{bhm} with thousands of latent variables. We address the controversial topic of the possible existence of a "magnitude step" \citep{Kelly_2010,Sullivan_2010} in type \Ia supernov\ae (\SNaeIa)---standardisable candles that enabled the discovery of the accelerated expansion of the Universe \citep{Perlmutter_1997,Perlmutter_1999,Riess_1998}---an intrinsic difference in magnitude correlated with the mass of their host galaxies, and its interplay with the host dust properties: see the references in \cite{Brout_Scolnic_2021} and \cite[hereafter \citetalias{Thorp_Mandel_2022}]{Thorp_Mandel_2022}, on which we base our modelling. So far, the problem has been plagued by an inability to treat all considered effects self-consistently due to the limitations of likelihood-based analyses, which are lifted by \gls{sbi}.

\section{Simulation-based model selection}

\paragraph{Bayesian model selection} assigns posterior probabilities \mpost to models $\Model[k] \in \qty{\Model[1], \mathellipsis, \Model[N]}$ (instead of to values of their parameters $\params_k$), conditional on observed data \data. The conventional approach is to compute the marginal likelihood (or \emph{evidence}) \evidence, which is the average likelihood \likelihood of parameters distributed according to the prior \prior:
\begin{equation}\label{eqn:mlike}
    \evidence = \int \likelihood\prior \dd{\params_k}
\end{equation}
(where the presence of \Model[k]'s parameters $\params_k$ implies conditioning on \Model[k] in the right-hand side). The prior belief in the model, \mprior, is then updated to its posterior probability in accordance with Bayes' theorem: $\mpost \propto \evidence \mprior$, normalised over all models considered.

As pointed out by \citet{Jeffrey_Wandelt_2023}, this has two disadvantages: first, it might be unclear what exactly the complete set of model parameters is, in what space they are defined (there are models with varying numbers of parameters: see e.g. trans-dimensional Monte Carlo \citep{transdim}), and what their likelihood is. For example, in cosmology, so-called \emph{selection effects} arise when the probability of detecting an object and including it in the analysed sample depends on the very parameters of interest. Even when the integral in \cref{eqn:mlike} is well defined, it is usually computationally prohibitive to evaluate for high-dimensional parameter spaces: variants of nested sampling, the \emph{de facto} standard technique for the task, typically only scale up to a few hundreds of parameters \citep[see \eg][for reviews]{Ashton_2022,Buchner_2023}, far from the millions required for contemporary cosmological data sets.\footnote{With the exception perhaps of proximal nested sampling, which scales to millions-dimensional models \citep{Cai_2022}.}

\paragraph{Marginal \acrlong{sbi}} circumvents both issues since the simulator abstracts latent parameters from the inference procedure altogether: latent stochastic variables sampled during a forward run are implicitly marginalised. For the purpose of Bayesian parameter estimation, the \gls{nn} can be trained to approximate either the likelihood, the posterior, or the likelihood-to-evidence ratio. The latter approach, called \gls{nre}, recasts the inference task into a classification problem between pairs $\params, \data \sim \joint$ versus $\params, \data \sim \prob(\params) \prob(\data)$ and uses the classification probability to derive the posterior over model parameters \params. \gls{nre} is founded on the well-known principle that, in order to minimise the Bayesian risk of misclassification, a classifier must base its decision on the ratio of the densities of the examples it has been trained on \citep[see \eg][]{Devroye_1996}, implying that if the classes represent data simulated according to the different models being compared (in proportion to the model priors \mprior), the \gls{nn} learns their posterior probabilities.

The ratio estimator used in \gls{nre} is usually trained to minimise the \gls{bce} loss \citep[see \eg][]{Hermans_2022} used for binary classification. In machine learning applications, the case is ubiquitously extended to multiple classes via the multi-class cross-entropy loss, whereby the \acrlong{nn} outputs one real number for each model considered: $\qty{x_1, \mathellipsis, x_N}$; these are then normalised via the softmax function: $y_k = \exp(x_k) / \sum_j \exp(x_j)$. Training a sufficiently expressive \acrlong{nn} to maximise the entry corresponding to the true model leads to it outputting (after the normalisation) the posterior probabilities of the models.

\paragraph{Related work.}
In the field of \SNIa analysis, \gls{sbi} has focused on marginal parameter inference: of cosmological parameters by using summary statistics derived in likelihood-based fits to light curves \citep{Weyant_2013,Alsing_Wandelt_2019,Wang_2022,Wang_2023b,Chen_2023,Karchev-sicret}, and of the properties of an individual object from its raw light curve \cite{Villar_2022}.

A number of studies have addressed simulation-based Bayesian model selection in general. \Citet{Jeffrey_Wandelt_2023} focused on loss functions for two-way model comparison with an emphasis on recovering accurate extreme Bayes factors.
\Citet{Radev_2021} proposed estimating a Dirichlet distribution over an arbitrary number of models using a \gls{nn} and variational optimisation, while \citet{Elsemuller_2023} advocated in favour of a cross-entropy loss, which we use in this work.


\section{Application to \SNIa analysis: magnitude step and dust laws}

\paragraph{The data} we analyse are light curves (collections of calibrated flux measurements in different passbands and at different times) of the \num{86} low-redshift \SNaeIa from the \gls{csp} \citep{Krisciunas_2017} previously investigated by \citetalias{Thorp_Mandel_2022} with a likelihood-based \gls{bhm}, which we re-implement as a forward simulator.
We fix the principal components of the \SNIa spectral time series to those inferred by \citet{Mandel_2022} and sample the remaining model parameters (including intrinsic light curve variations and dust optical depth) from their respective priors (the \gls{nn} implicitly marginalises them for the purposes of model comparison) and implement \num{6} models whose posterior probabilities we wish to evaluate:
\begin{itemize}[leftmargin=1.5em, label=\textbullet]
    \item the possible existence of an intrinsic magnitude step between \SNae in low- and high-mass hosts,
    such that the magnitude scatter follows $\Normal(\stepmag, \sigmares^2)$ for \SNae in  high-mass galaxies ($\log_{10}(M_*/M_\odot) > \num{10.5}$), and $\Normal(0, \sigmares^2)$ otherwise\footnote{Our \stepmag has opposite sign to \citetalias{Thorp_Mandel_2022}: here, \stepmag $<0$ corresponds to brighter \SNaeIa in more massive hosts.}; we place a uniform hyperprior $\stepmag \sim \Uniform(\num{-0.2}, \num{0.2})$ and a broad hyperprior on \sigmares as in \citet{Mandel_2022}); we label with "\texttt{dM}" ("\texttt{M0}") the model with (without) a magnitude step, so that \texttt{dM} \textrightarrow{} \texttt{M0} when $\stepmag \rightarrow 0$;
    
    \item the population of host dust parameters \Rvs (which in all cases are restricted to the range $\interval{\num{0.5}}{\num{6}}$ as in \citetalias{Thorp_Mandel_2022}), describing the wavelength dependency of dust absorption in the Fitzpatrick law \citep{Fitzpatrick_1999}:
    \begin{itemize}[leftmargin=1.5em, label=\textbullet]
        \item a "global" dust model, the simplest among them, has $\Rvs = \meanRv$, \ie all \SNae are subject to the same dust law (albeit with individual optical depth described by $A_V^s$);
        \item a "local" dust model, assuming a hierarchical relationship $\Rvs \sim \Normal(\mu_R, \sigmaRv^2)$, \ie a single population of dust; "local" \textrightarrow{} "global" when $\sigmaRv \rightarrow 0$; 
        \item a "split" dust model, with two independent distributions of \Rvs for high- and low-mass hosts: $\Rvs \sim \Normal(\meanRv^{\text{low}}, (\sigmaRv^{\text{low}})^2) \text{ or } \Normal(\meanRv^{\text{high}}, (\sigmaRv^{\text{high}})^2)$, such that "split" \textrightarrow{} "local" when $(\cdot)^{\text{low}} \rightarrow (\cdot)^{\text{high}}$.
    \end{itemize}
    We use the same priors on global dust parameters
    as in \citetalias{Thorp_Mandel_2022}, a fixed mass split location at $\log_{10}(M_*/M_\odot) = \num{10.5}$ (resulting in a \num{49}/\num{37} split), and stellar masses $M_*$ as included in \SNANA \citep{SNANA}, ignoring stellar mass uncertainty (see \cite{Shariff_2016}).
\end{itemize}
We fix the \SN cosmological redshifts (after peculiar velocity corrections \citep{Carrick_2015}) and the cosmological model to that used in \citetalias{Thorp_Mandel_2022}: a flat $\Lambda$CDM with $\Omega_{\text{m}0} = \num{0.28}$ and $H_0 = \SI{73.24}{\km / \s / Mpc}$ (and \SNIa absolute magnitude $M_0 = \num{-19.5}$).
Overall, our models have \num{47} parameters \emph{per \SN} (\num{42} of them describing the residual correlated light curve variability) for a total of more than \num{4000} for the analysed data set with \num{86} \SNaeIa. For comparison, current state-of-the-art compilations of about \num{2000} \SNaeIa \citep{Scolnic_2022} would require \num{\sim e5} latent variables.

\begin{wrapfigure}{l}{0.42\linewidth}%
    \includegraphics{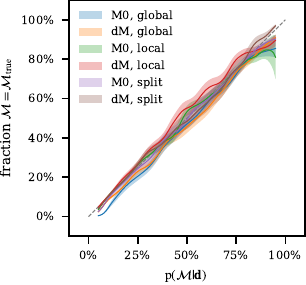}%
    \vspace{1em}%
    \caption{Reliability diagram for the trained classifier network.\label{fig:validation}}
\end{wrapfigure}
\textbf{The \acrlong{nn}} we use is implemented in \texttt{pytorch} \citep{pytorch} and is based solely on fully connected layers. Full details about the architecture and training are given in \cref{sec:nn}. Before showing results on the real data set, we validate the performance of the trained classifier using a set $\qty{\data_i}$ simulated from models $\qty{\Model[i]}$ in proportion to the prior model probabilities (in this case, in equal amounts).

\textbf{Calibration.} We first plot the "reliability diagram"\footnote{\Citet{Jeffrey_Wandelt_2023} use the same diagnostic, calling it "blind coverage testing".} \citep{DeGroot_Fienberg_1983}, which shows the fraction of examples that were simulated from a given model versus the posterior probability of that model given the simulated data. To produce \cref{fig:validation} we bin the validation examples according to the network output for model $k$ (\ie the posterior probability $\prob(\Model[k]\given\data_i)$) and within each bin calculate the fraction of examples that were actually simulated from model $k$.
The nearly diagonal lines we observe indicate good calibration.

\begin{figure}%
    \centering%
    \vphantom{ }  
    \clap{\includegraphics[trim={0.4803in 0 0.3611in 0}]{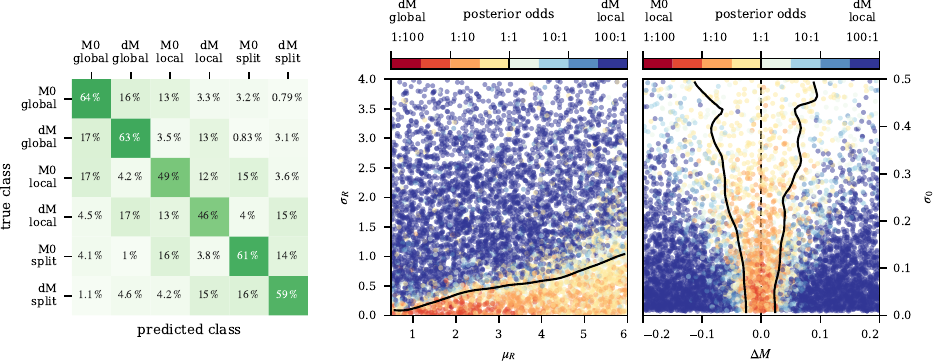}}%
    \caption{Evaluation of the trained classifier network on the validation set of simulations. 
    \emph{Left:} Each row shows the posterior over models (as labelled above), averaged over a collection of data simulated with the model indicated on the left.
    \emph{Right:} $\log_{10}$ Bayes factors (evidence ratios) for different simulated datasets as a function of the input parameters. For the \meanRv--\sigmaRv plot the compared models are \enquote{local} and \enquote{global} ($\sigmaRv = 0$), marginalising over \stepmag, while for the \stepmag--\sigmares plot, the models are \enquote{\texttt{dM}} and \enquote{\texttt{M0}} ($\stepmag = 0$), assuming a non-split \Rvs distribution (\enquote{local}). The solid black lines indicate parameters leading \emph{on average} to equal posterior odds.
    \label{fig:confusions}}
\end{figure}

\textbf{Refinedness} (in the sense of \cite{DeGroot_Fienberg_1983}).
We show in \cref{fig:confusions} (left) the average posterior probabilities for data simulated with a given model. A refined classifier would assign the most probability to the "correct" model, leading to a pronounced diagonal; but unlike in usual machine learning applications, Bayesian model comparison assigns non-zero posterior probability to all models (\ie non-zero off-diagonal entries). The prominence of the diagonal, then, depends both on how powerful the data itself is in distinguishing the models as well as on the parameters' priors \citep{Trotta2008}.

Owing to amortisation, we are able to explore Bayes factors (ratios of evidences) across a range of ground-truth parameters of simulated data, which is computationally unfeasible with traditional methods. \Cref{fig:confusions} (right), which compares nested models ("local" \textrightarrow{} "global" in \meanRv--\sigmaRv space and \texttt{dM} \textrightarrow{} \texttt{M0} in \stepmag--\sigmares space), clearly demonstrates Occam's razor: data resulting from parameters sufficiently close to the location of the nested model ($\sigmaRv=0$ or $\stepmag=0$) favour the simpler model (yellow/red regions). We also observe that, naturally, a step in magnitudes is harder to detect when their scatter (\sigmares) is larger. A scatter in \Rvs (\ie $\sigmaRv > 0$) is also harder to detect when \meanRv is large because, in that region, the effect on data is smaller due to the non-linear nature of the dust law.

\begin{figure}%
    \centering%
    \vphantom{ }  
    \clap{\includegraphics[trim={0.1689in 0 0.3611in 0}]{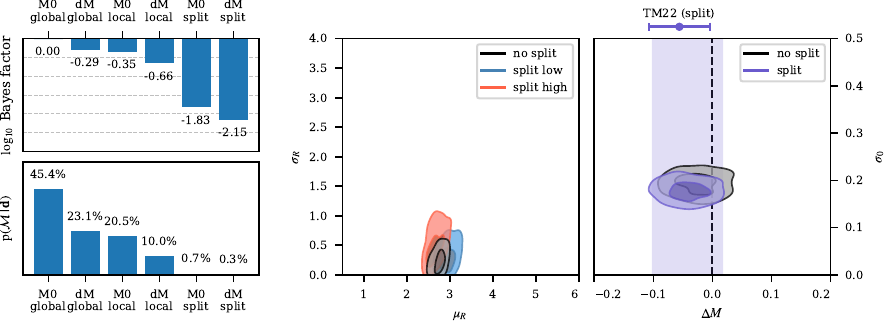}}%
    \caption{Posteriors from \acrshort{csp} data.
    \emph{Left:} Models' posterior probabilities (bottom) and (top) $\log_{10}$ Bayes factor with respect to the highest-ranked model (\texttt{M0}, global: no step, global dust law).
    \emph{Right:} Approximate marginal posteriors ($1\sigma$ and $2\sigma$) from \acrshort{nre}. The \meanRv--\sigmaRv plot compares the posteriors for ($\cdot^{\text{low}}$, $\cdot^{\text{high}}$) from the \enquote{split} model, with the result for a single dust-law distribution (\enquote{no split}). The \stepmag--\sigmares plot compares posteriors from the same two models. The shaded strip is the 1-D $2\sigma$ marginal \stepmag posterior from the \enquote{split} model, in agreement with \citetalias{Thorp_Mandel_2022} ($2\sigma$ error bar above).
    \label{fig:posteriors}}
\end{figure}

\paragraph{Results} from the \gls{csp} data set are presented in \cref{fig:posteriors} in terms of posterior model probabilities and Bayes factors with respect to the most probable model: a global dust law and no magnitude step. Our results follow Occam's razor, with no clear preference for a mass step and a mildly disfavoured (by a factor \num{\approx 2}) spread of \Rvs. A split in the dust laws for low- and high-mass hosts is clearly disfavoured, regardless of the magnitude step, with a Bayes factor of \num{\approx 100}, contrary to the conclusions of both \citet{Thorp_Mandel_2022} and \citet{Brout_Scolnic_2021}.

In \cref{fig:posteriors}, we also present posteriors (derived via \gls{nre} trained on the same simulations used for the model comparison network), which support the conclusions of model comparison. In agreement with \citetalias{Thorp_Mandel_2022}, we find a magnitude step of $\stepmag \approx \num{-0.05}$, and approximately $2\sigma$ away from \num{0}, with the results only mildly affected by the dust model. We find a larger value $\sigmares \approx \num{0.2}$ (\cf \num{\approx 0.1} in \citetalias{Thorp_Mandel_2022})
since this quantity in our analysis absorbs all residual variability present in the data, including peculiar velocity uncertainties, which we do not model explicitly.
All of the global dust parameter posteriors are in good agreement with \citetalias[fig.~8]{Thorp_Mandel_2022}, and we obtain similar posteriors when treating low- and high-mass hosts separately as when we assume a single dust distribution (after marginalising over \stepmag in all cases). This justifies the "split" dust model being strongly disfavoured, due to its larger prior volume.

\section{Conclusions}

Enabled by neural \gls{sbi}, we have performed Bayesian model comparison on an unsolved problem in cosmology that requires realistic modelling of \SNIa light curves and marginalising over thousands of latent variables.
A demonstration of Occam's razor, our results from low-redshift \SNIa data favour a global dust law and no magnitude step (with \SI{45}{\percent} posterior probability up from \SI{16.6}{\percent} \emph{a priori}). The existence of a magnitude step or a distribution of \Rvs remain plausible (with posterior odds of $\approx 1:2$), while a split in global dust populations across $\log_{10} M_*/M_\odot = \num{10.5}$ is disfavoured with odds in excess of 100:1. We emphasise, however, that Bayesian model comparison is always dependent on the prior volumes considered. The scalability of our approach allows it to be applied to much larger data sets than demonstrated here, both present and future, with even more sophisticated Bayesian models (\eg marginalising out the location of the mass split, accounting for stellar mass uncertainty), and more realistic simulators (self-consistently estimating redshifts and peculiar velocities, including selection effects and non-\Ia contamination), ushering in the era of principled simulation-based fully Bayesian \SNIa cosmology.

\begin{ack}
RT acknowledges co-funding from Next Generation EU, in the context of the National Recovery and Resilience Plan, Investment PE1 -- Project FAIR ``Future Artificial Intelligence Research''. This resource was co-financed by the Next Generation EU [DM 1555 del 11.10.22]. RT is partially supported by the Fondazione ICSC, Spoke 3 ``Astrophysics and Cosmos Observations'', Piano Nazionale di Ripresa e Resilienza Project ID CN00000013 ``Italian Research Center on High-Performance Computing, Big Data and Quantum Computing'' funded by MUR Missione 4 Componente 2 Investimento 1.4: Potenziamento strutture di ricerca e creazione di ``campioni nazionali di R\&S (M4C2-19)'' -- Next Generation EU (NGEU).
CW acknowledges funding from the European Research Council (ERC) under the European Union’s Horizon 2020 research and innovation programme (Grant agreement No. 864035).
\end{ack}


\begingroup
\footnotesize
\setlength{\bibsep}{3pt}
\bibliography{simsims,simsims-extra}
\endgroup


\clearpage
\begin{appendix}
    \section{Neural network architecture and training}\label{sec:nn}

    We use a \acrlong{nn} that consists entirely of fully connected linear layers followed by online whitening (which shifts and rescales its inputs to have null mean and unit standard deviation) and \gls{relu} non-linearities. Since the number of observations for each supernova varies from one object to another, we use a bespoke linear layer $\mathbb{R}^{\nsnobs} \rightarrow \mathbb{R}^{256}$ to embed each \SN in a common-dimensional space. The embedding is processed by two more layers (shared among all \SNae), resulting in a \SN featurisation in $\mathbb{R}^{32}$. The resulting $\nsn=\num{86}$ feature vectors are flattened to form a $\num{86} \times \num{32} = 2752$-dimensional representation of the whole data set, which is fed through three additional layers leading finally to the \num{6} predicted class probabilities (unnormalised logits input into a \texttt{CrossEntropyLoss}). We implement the network (detailed in \cref{tbl:nn}) in \texttt{pytorch} \citep{pytorch}, using a \SI{50}{\percent} dropout \citep{dropout} after the flattening layer. We train on a single Nvidia A-100 \gls{gpu} using \num{96 000} examples from each of the \num{6} models (set to fit in the \gls{gpu} memory) and a \texttt{OneCycle} learning rate schedule \citep{onecyclelr}. Generating the training data and training until convergence (for about \num{100000} steps) took about \SI{1}{\hour} each.
    
    \begin{table}[h!]
        \caption{Architecture of the neural network we use.\label{tbl:nn}}
        \centering
        \begin{tabular}{r l l l}
            \toprule
            
            input & shape: $(\sum_{s=1}^{\nsn} \nsnobs, )$ & $\WhitenOnline()$ \\ \midrule
            loop $s \in 1,\mathellipsis,\nsn$ & $\Linear(\nsnobs, 256)$ & $\WhitenOnline()$ & $\ReLU()$ \\[0.5em]
            Stack & \multicolumn{2}{l}{shape: $\nsn \times (256,) \rightarrow (\nsn, 256)$} \\ \midrule
            
            \multirow{3}{*}{\shortstack[r]{\SNae as\\batch dim.}}
            & $\Linear(256, 256)$ & $\WhitenOnline()$ & $\ReLU()$ \\ 
            & $\Linear(256, 256)$ & $\WhitenOnline()$ & $\ReLU()$ \\ 
            & $\Linear(256, 32)$ \\[0.5em]
            Flatten & \multicolumn{2}{l}{shape: $(\nsn, 32) \rightarrow (\nsn \times 32,)$}
            & $\Dropout(0.5)$ \\ \midrule
            
            & $\Linear(256, 256)$ & $\WhitenOnline()$ & $\ReLU()$ \\ 
            & $\Linear(256, 256)$ & $\WhitenOnline()$ & $\ReLU()$ \\ 
            & $\Linear(256, 256)$ \\ \midrule
            & $\Linear(256, 6)$ & classifier output & ($\rightarrow \CrossEntropyLoss$) \\
            
            \bottomrule
        \end{tabular}
    \end{table}
\end{appendix}

\end{document}